\documentclass[runningheads]{llncs}

\usepackage[utf8]{inputenc}
\usepackage[hidelinks]{hyperref}
\usepackage{graphicx}
\usepackage{url}
\usepackage{subcaption}
\usepackage{verbatimbox}
\usepackage{amssymb}
\usepackage{amsmath}

\begin{document}

\mainmatter

\title{Advanced Memory Buoyancy\\for Forgetful Information Systems}

\author{
  Christian Jilek\inst{1,2} \and
  Jessica Chwalek\inst{2} \and
  Sven Schwarz\inst{1} \and\\
  Markus Schröder\inst{1,2} \and
  Heiko Maus\inst{1} \and
  Andreas Dengel\inst{1,2}
}

\authorrunning{C. Jilek et al.}

\institute{
  Smart Data \& Knowledge Services Department, DFKI GmbH,\\Kaiserslautern, Germany\\
  \email{\{christian.jilek,sven.schwarz,markus.schroeder,\\heiko.maus,andreas.dengel\}@dfki.de}\\ \and
  Computer Science Department, TU Kaiserslautern, Kaiserslautern, Germany\\
  \email{j\_chwalek14@cs.uni-kl.de}
}

\maketitle

\begin{abstract}
Knowledge workers face an ever increasing flood of information in their daily lives.
To counter this and provide better support for information management and knowledge work in general, we have been investigating solutions inspired by human forgetting since 2013.
These solutions are based on \textit{Semantic Desktop} (SD) and \textit{Managed Forgetting} (MF) technology.
A key concept of the latter is the so-called \textit{Memory Buoyancy} (MB), which is intended to represent an information item's current value for the user and allows to employ forgetting mechanisms.
The SD thus continuously performs information value assessment updating MB and triggering respective MF measures.
We extended an SD-based organizational memory system, which we have been using in daily work for over seven years now, with MF mechanisms directly embedding them in daily activities, too, and enabling us to test and optimize them in real-world scenarios.

In this paper, we first present our initial version of MB and discuss success and failure stories we have been experiencing with it during three years of practical usage.
We learned from cognitive psychology that our previous research on \textit{context} can be beneficial for MF.
Thus, we created an advanced MB version especially taking user context, and in particular context switches, into account.
These enhancements as well as a first prototypical implementation are presented, too.

\end{abstract}

\section{Introduction}
\label{sec:introduction}
Knowledge workers face an ever increasing flood of information in their daily lives, which is even intensified by technological trends like digital transformation.
To counter this and provide better support for information management and knowledge work in general, we have been investigating solutions inspired by human forgetting since 2013, starting with the EU-project
\textit{ForgetIT}\footnote{2013--2016, \url{www.forgetit-project.eu}}
and continuing in the
\textit{Managed Forgetting} project\footnote{2016--2019, \url{www.spp1921.de/projekte/p4.html.de}},
which is part of the recent priority program on ``Intentional Forgetting in Organizations''  by the German Research Foundation (DFG).
On the one hand, these solutions are based on semantic technologies, especially semantic graphs/networks, in a knowledge work scenario consisting of a \textit{Semantic Desktop} (SD) ecosystem for individual users embedded in an evolutionary organizational memory for groups.
On the other hand, they are based on \textit{Managed Forgetting} (MF) by which we understand an escalating set of measures ranging from temporal hiding to condensation to deletion, thus overcoming the binary keep-or-delete paradigm.
A key concept of MF is the so-called \textit{Memory Buoyancy} (MB), which is intended to represent an information item's current value for the user and allows to employ forgetting mechanisms.
Following the eat-your-own-dog-food credo, we extended our SD-based organizational memory system, which we have been using in daily work for over seven years now, with MF mechanisms.
These new forgetting capabilities were thus directly embedded in daily activities, too, enabling us to test and optimize them in real-world scenarios.
We learned from cognitive psychology that our previous research on \textit{context} can be beneficial for MF.
Thus, a major enhancement we incorporated into a successor of the initial MB was to take user context into account.

In this paper, we first provide more details about SD and MF (Section \ref{sec:sdmf}).
Next, we present the initial MB and discuss success and failure stories we have been experiencing with it during three years of practical usage (Section \ref{sec:mb}).
In Section \ref{sec:advmb}, we introduce our advanced MB version, which takes user context, and in particular context switches, into account.
Last, we conclude this paper and give an outlook on possible future work in Section \ref{sec:conclusion}.
\section{Semantic Desktop \& Managed Forgetting}
\label{sec:sdmf}

\textbf{Semantic Desktop.}
The \textit{Semantic Desktop} (SD) \cite{sauermann2005semdesk} is especially intended to capture knowledge that emerges from individuals and then spreads into groups like project teams.
SD brings \textit{Semantic Web}\footnote{\url{www.w3.org/standards/semanticweb}} technology to users' computing devices using a knowledge representation, i.e. giving resources unique identifiers (URIs) and allowing to make statements about them, e.g. using RDF\footnote{\url{www.w3.org/RDF}}, resulting in a semantic graph.
Information items (files, mails, contacts, events, topics, \ldots) that are separated on the computer (file system, mail client, web browser, \ldots) but are related to each other in a person's mind, can thus be semantically represented and interlinked in a machine understandable way.
As soon as such an item is semantically represented, it is called a ``thing'', which describes the item uniquely as an URI complemented by further statements like its type or a reference to the originating resource such as an URL or message-id of an e-mail.
Capturing a user's mental model as accurate as possible is done in a \textit{Personal Information Model} (PIMO) \cite{SauermannVanElstDengel2007}, which serves as the basis for knowledge representation in the SD.
Shared parts of multiple PIMOs result in a \textit{Group Information Model} (GIMO) forming the basis for an organizational memory.
\\

\begin{figure}[h]
  \centering
  \includegraphics[width=1\columnwidth]{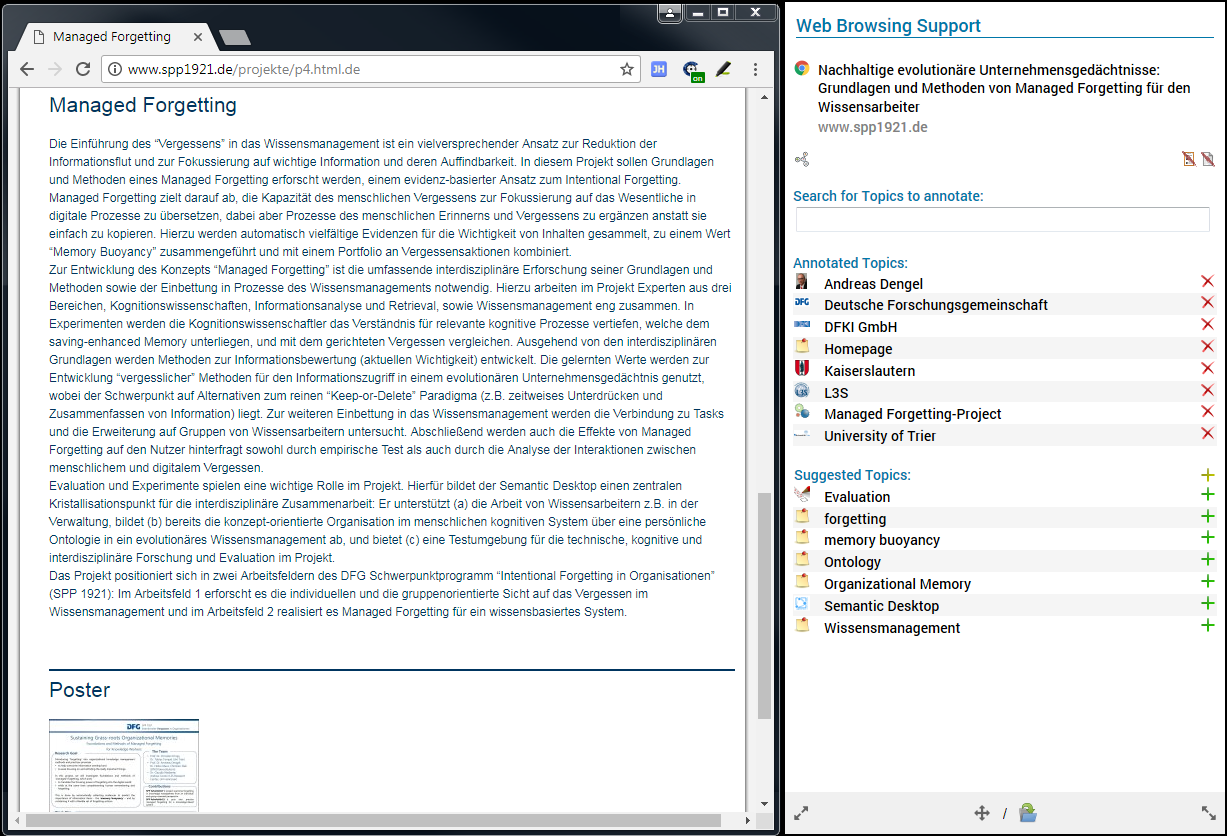}
  \caption{SD sidebar (right) supporting user activities like web browsing (left)}
  \label{fig:sidebar}
\end{figure}

\begin{figure}[h!]
  \centering
  \includegraphics[width=0.88\columnwidth]{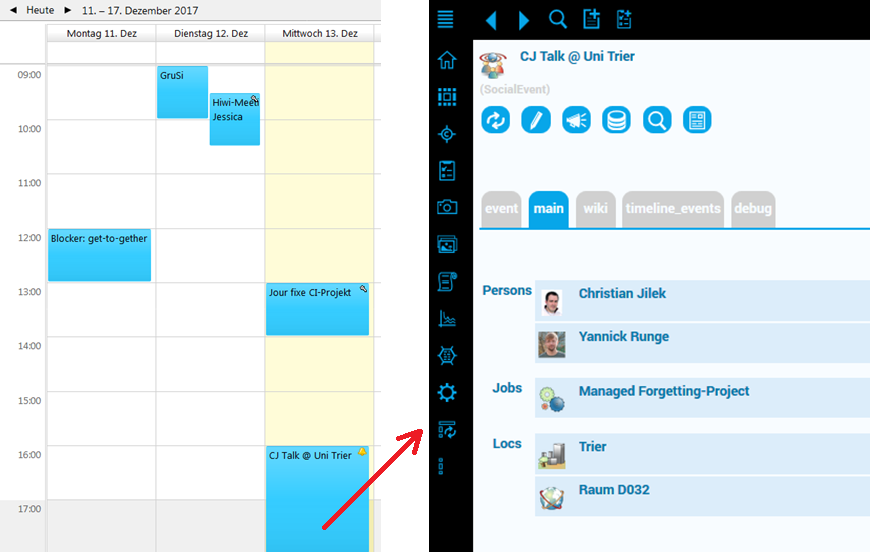}
  \caption{Browsing the representation of a particular calendar event in PIMO}
  \label{fig:pimodav}
\end{figure}

\begin{figure}[h!]
  \centering
  \includegraphics[width=0.88\columnwidth]{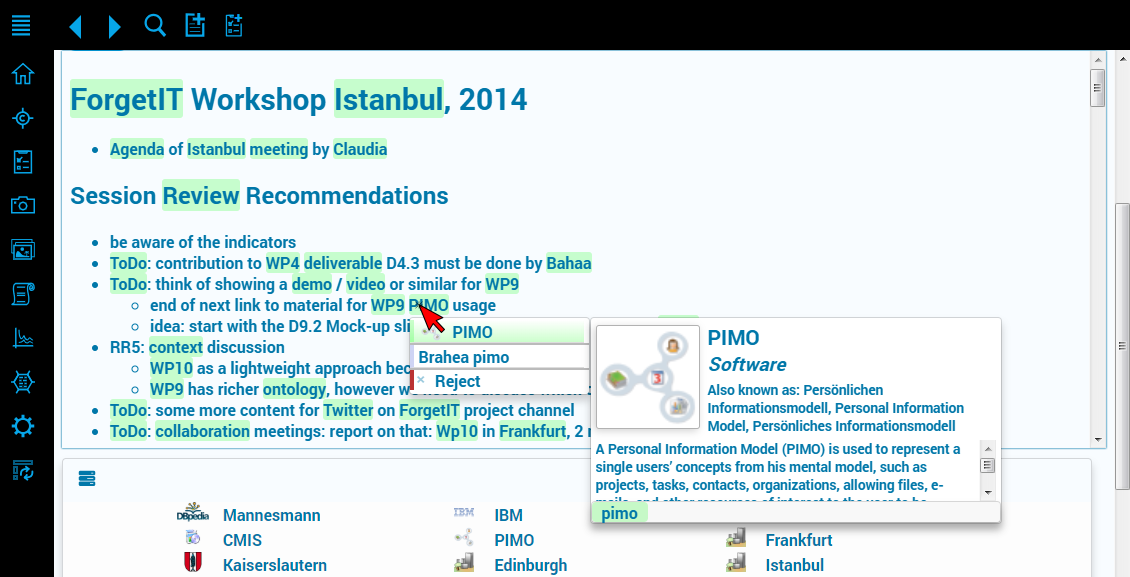}
  \caption{Semantic Text Editor \textit{Seed}: data contextualization while typing}
  \label{fig:seed}
\end{figure}

\begin{figure}[t]
  \centering
  \includegraphics[width=1\columnwidth]{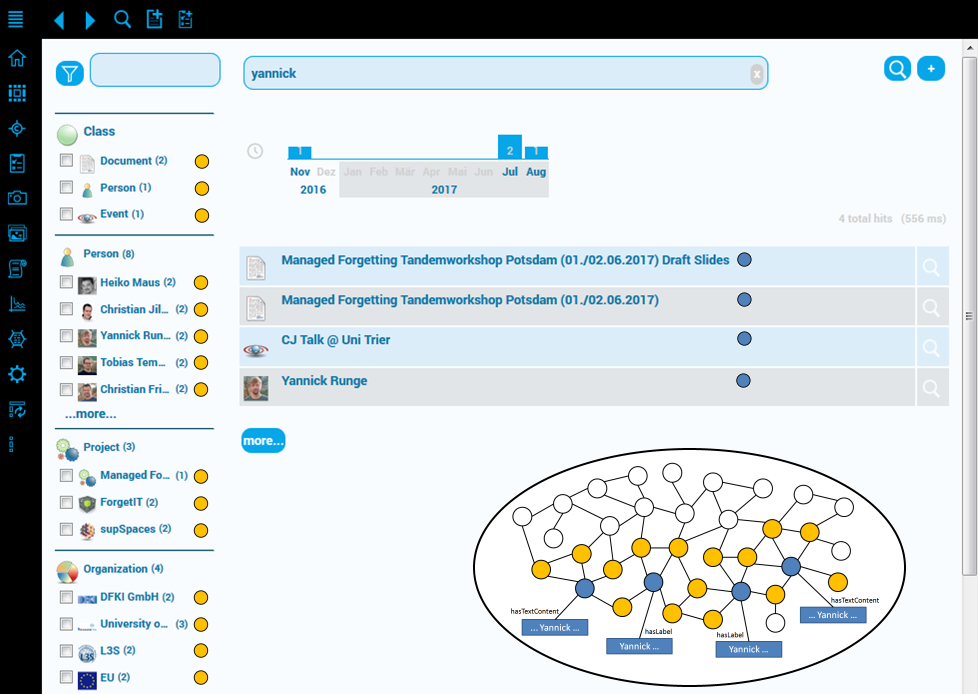}
  \caption{Using dynamic facets (yellow) to search things in PIMO (blue)}
  \label{fig:facetedsearch}
\end{figure}

\noindent
\textbf{SD Prototype.}
Our current SD research prototype \cite{MausSchwarzDengel2013} uses micro plug-ins \cite{JilekSchroederSchwarz+2018} to capture in-app events and performs information extraction on these evidences in order to elicit the current user activity/context.
Then, SD decides on and executes appropriate support measures.
Users interact with the system using a sidebar and a web app called \textit{PIMO5} (referring to \textit{HTML5} as its base technology).
Figure \ref{fig:sidebar} shows a user browsing a website while the SD sidebar analyzes the visited website in real-time \cite{JilekSchroederNovik+2018NER} showing concepts of the user's PIMO that are mentioned on that particular site.
Users may thus easily tag this page with the respective topics.
Similar plug-ins are available for email clients or the file system.
SD's WebDAV%
\footnote{\url{https://tools.ietf.org/html/rfc4918}}-based calendar \textit{PimoDAV} is depicted in Figure \ref{fig:pimodav}:
events entered into PimoDAV are automatically analyzed and added to the user's PIMO.
Writing textual annotations (comments, notes, etc.) is possible using \textit{Seed} \cite{EldesoukyMausSchwarz+2015}, a semantic text editor (see Figure \ref{fig:seed}).
While typing, Seed scans the text for entities found in the user's PIMO or other sources like DBpedia%
\footnote{a semantic representation of Wikipedia, \url{https://wiki.dbpedia.org/}}.
In the screenshot, those entities are highlighted in green.
When hovering over such a concept further details appear (as it is the case for \textit{PIMO (software)} in the screenshot).
Content is thus contextualized while typing easing later retrieval.
Speaking of retrieval: a user's PIMO can be searched using dynamically generated facets like illustrated in Figure \ref{fig:facetedsearch}.
Here, the keyword \textit{yannick} (first name of a colleague) is entered.
The PIMO is then searched for all things having a literal (label, comment, etc.) matching that particular keyword.
The resulting list (highlighted using blue dots) can be further filtered using facets dynamically calculated from the results' relations in the semantic graph, e.g. types, associated persons, projects, locations, etc. (highlighted with yellow dots).

So far, the system would sooner or later run into scaling problems typically found in SD systems \cite{DraganDecker2012}.
That is another reason why we started investigating solutions based on human forgetting, which are discussed in the following.
\\

\noindent
\textbf{Managed Forgetting.}
We started our investigations on information management and knowledge work support measures based on human forgetting and SD technology in the previously mentioned \textit{ForgetIT} project.
It was an EU-funded project consisting of a large international team.
In its course, the \textit{Memory Buoyancy} (MB) representing an information item's short-term value for the user as well as the \textit{Preservation Value} (PV) as its long-term perspective counterpart were conceived \cite{NiedereeKanhabuaGallo+15,NiedereeKanhabuaTran+2018}.
A second aspect of the project was contextualized remembering:
here, the idea is to store contextual metadata in addition to the actual material to ensure future comprehension of the text, e.g. who were the mentioned persons, terms, topics, etc.
An application in this regard is the aforementioned semantic text editor \textit{Seed} \cite{EldesoukyMausSchwarz+2015} that allows for data contextualization while typing.
Last not least, the project introduced the concept of \textit{Managed Forgetting} (MF), by which we understand an escalating set of measures overcoming the binary keep-or-delete paradigm.
They range from temporal hiding, to condensation, to adaptive synchronization, archiving and deletion.
As stated before, MB is a key concept of MF and will be presented more thoroughly in the next section.
Additionally, we show examples of how we incorporated MF into the SD.
Further details going beyond this short overview can be found in \cite{MausJilekSchwarz2018}, \cite{JilekSchwarzMaus+2016} and \cite{JilekMausSchwarz+2015}.
Focusing more on self-(re)organization aspects realized by means of MF, the more recent and still ongoing \textit{Managed Forgetting} project refines several outcomes of ForgetIT like temporal hiding or MB.
The extensions of the latter are especially addressed in Section \ref{sec:advmb}.

\section{Memory Buoyancy}
\label{sec:mb}

\noindent
\textbf{Memory Buoyancy.}
Memory Buoyancy (MB) \cite{NiedereeKanhabuaGallo+15,NiedereeKanhabuaTran+2018} is intended to represent an information item's current (short-term) value for the user.
It follows the metaphor that items starting to lose relevance ``sink away'' from the user's sight, but are brought back up (by their higher buoyancy) as soon as they regain relevance.

In this section, we present our initial version of MB as well as success and failure stories we experienced with it during three years of daily usage in our organizational memory system.
Last, we discuss problems leading to these failures and how we solved or intend to solve them.
\\

\noindent
\textbf{Related Work.}
MB is a form of information value assessment (IVA), to our best knowledge, the only one designed and implemented according to findings of cognitive psychology about human memory and cognition.
However, there are also other approaches mentioned in literature, which we will shortly mention in the following without going into details:
Wijnhoven, Amrit et al. investigated identifying and dealing with \textit{information waste}, for example in the file system \cite{WijnhovenAmritDietz2014} or on the internet \cite{AmritWijnhovenBeckers2015}.
Hasan \& Burns speak of \textit{waste data}:
In \cite{HasanBurns2013}, they analyze the impact of such data on computing environments and propose solutions on how to cope with it.
Turczyk et al. \cite{TurczykGroeplLiebau+2007} investigated \textit{file valuation} in the area of information lifecycle management, which seeks to store files on different storage systems according to their (business) value.
Gyllstrom \& Pedersen propose \textit{LostRank} \cite{GyllstromPedersen2010}, an approach to estimate which documents are most likely to be lost for the user (e.g. important but not recently used) and are thus harder to re-find.
\\

\begin{figure}[h]
  \centering
  \begin{subfigure}[t]{0.48\textwidth}
    \includegraphics[width=\linewidth]{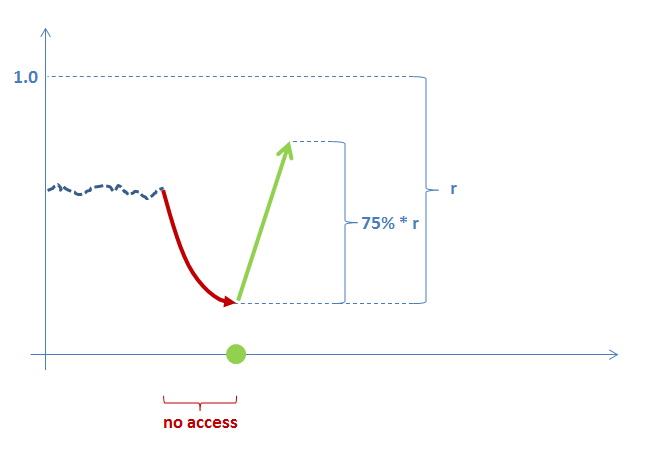}
    \caption{single access}
    \label{fig:mbdesign1}
  \end{subfigure}
  \hfill
  \begin{subfigure}[t]{0.48\textwidth}
    \includegraphics[width=\linewidth]{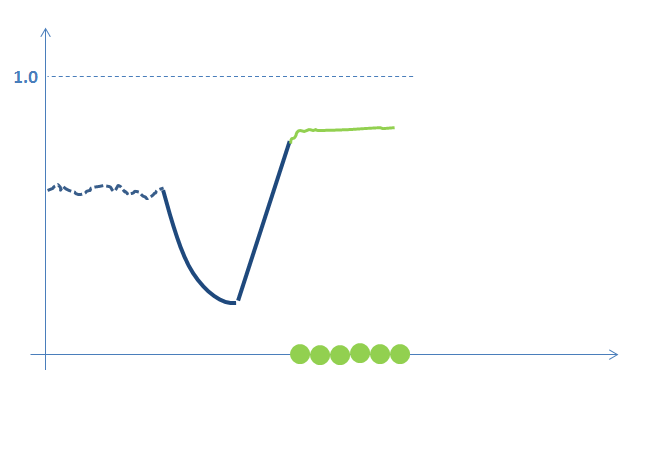}
    \caption{multiple accesses in quick succession}
    \label{fig:mbdesign2}
  \end{subfigure}
  \\
  \begin{subfigure}[t]{0.48\textwidth}
    \includegraphics[width=\linewidth]{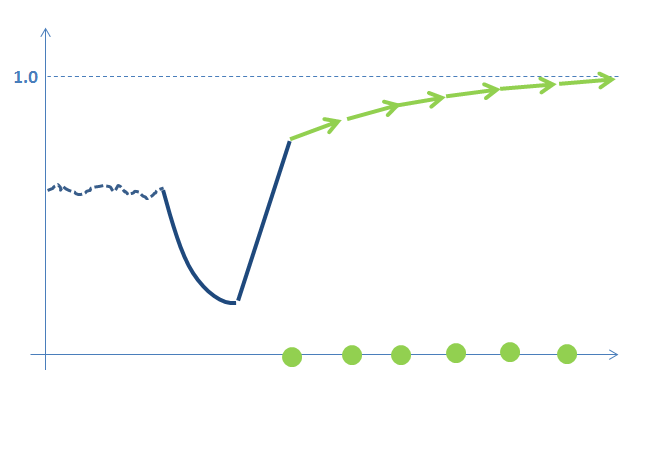}
    \caption{multiple accesses every day}
    \label{fig:mbdesign3}
  \end{subfigure}
  \hfill
  \begin{subfigure}[t]{0.48\textwidth}
    \includegraphics[width=\linewidth]{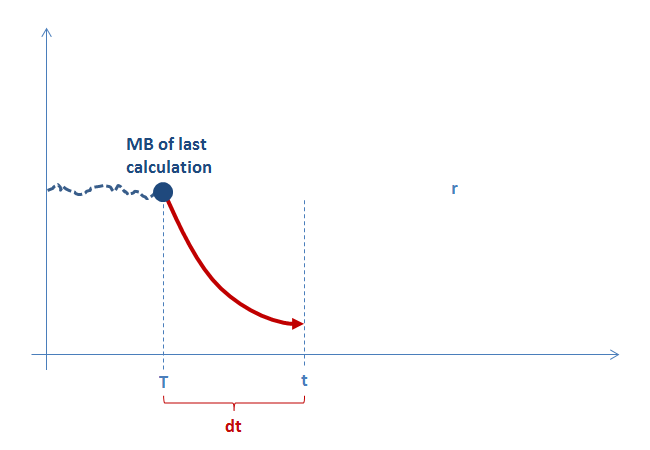}
    \caption{static and dynamic parts of MB}
    \label{fig:mbdesign4}
  \end{subfigure}
  \caption{Selected design principles for MB in a SD}
\end{figure}

\noindent
\textbf{Design Principles of MB in SD.}
The calculation of MB in our SD evolved from the insights of the approach in \cite{TranSchwarzNiederee+2016} and finally follows design principles shortly presented in the following.
The basic ones are inspired by human brain activity applied to the user's mental model as represented in a semantic graph and discussions with the team of Prof. Logie (Psychology, University of Edinburgh) who presented their insights in \cite{LogieWoltersNiven2018}.
\begin{itemize}
\item MB is updated every time a thing is stimulated (which in worst case could be after each click of a user).
\item MB depends on:
  \begin{itemize}
  \item user interaction (e.g. viewing, modifying, annotating),
  \item the thing itself (e.g. mails are assumed to be more ephemeral than a presentation), and
  \item its connections in the semantic network (PIMO).
  \end{itemize}
\item MB values are normalized ($0 \leq \text{MB} \leq 1$).
\item Single access of an item should not directly lead to an MB of 1.0 (Fig. \ref{fig:mbdesign1}).
\item Multiple accesses in quick succession (e.g. every minute) are treated reluctantly (Fig. \ref{fig:mbdesign2}).
\item Multiple accesses every day, for example, will saturate against 1.0 (Fig. \ref{fig:mbdesign3}).
\item MB drops for things that are not stimulated:
  \begin{itemize}
  \item first, there is a steep decline, then
  \item a long-tail of slow decline.
  \end{itemize}
\item We added rules and heuristics to deal with requirements of various domains:
  \begin{itemize}
  \item upcoming events should stimulate connected things,
  \item finished items (tasks, events) shall decrease faster unless other indicators speak against this,
  \item times with low user interaction should not lead to massive decay in MB
  \end{itemize}
\item MB utilizes the semantic network (PIMO): a form of \textit{spreading activation} \cite{Crestani1997} is performed using additional heuristics for
  \begin{itemize}
  \item types (e.g. emails are faster forgotten than persons), or
  \item the number of relations that connect things.
  \end{itemize}
\end{itemize}
In addition to these conceptional design principles, there are also technical requirements:
\begin{itemize}
\item MB is time-dependent which poses a challenge for calculation as MB values change permanently just as time passes.
\item MB values must always be up-to-date (simply doing an overnight update is not satisfactory).
\item Especially since it is a very interactive system, its high overall performance must be kept despite the additional MB calculation.
\end{itemize}
We thus split MB into two parts: a static, time-independent part and a dynamic, time-dependent one as shown in Figure \ref{fig:mbdesign4}.
This allows us to only update MB when applications or services actually query the resources, which reduces the additional calculation effort.

As mentioned before, this is only a short overview of the design principles: more explanations and details can be found in \cite{MausJilekSchwarz2018}.
\\

\noindent
\textbf{Success Stories.}
We incorporated MB in our productively used SD-based organizational memory system.
Thus, we experienced several success and failure stories during three years of practical usage.
In this section, we present one prominent example of each category.

\begin{figure}[h!]
  \centering
  \includegraphics[width=1\columnwidth]{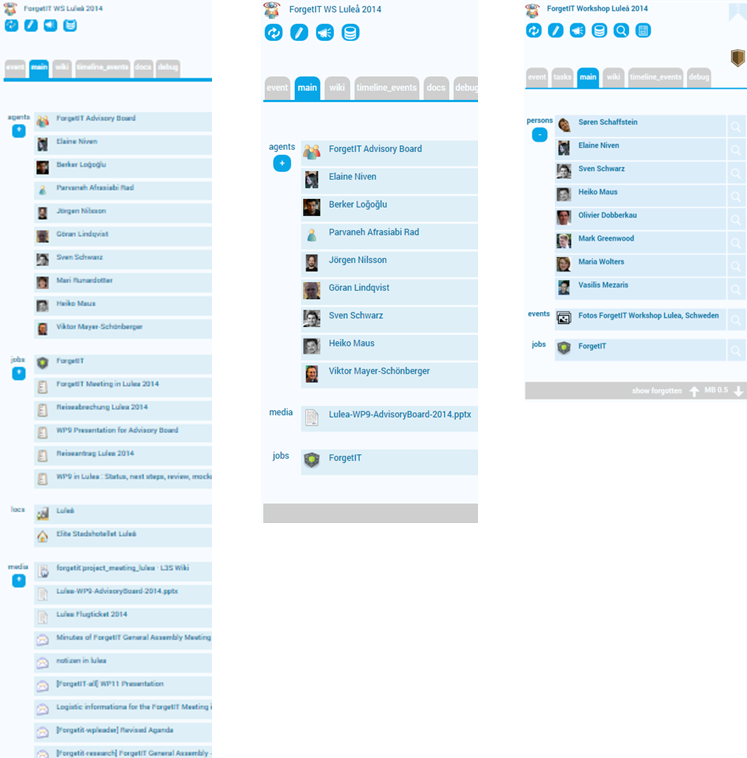}
  \caption{Details of a calendar event gradually fading out: a comparison of the state during the meeting (left), after 8 months (middle), and after 2 years (right)}
  \label{fig:fadingout}
\end{figure}

\begin{figure}[h!]
  \centering
  \includegraphics[width=1\columnwidth]{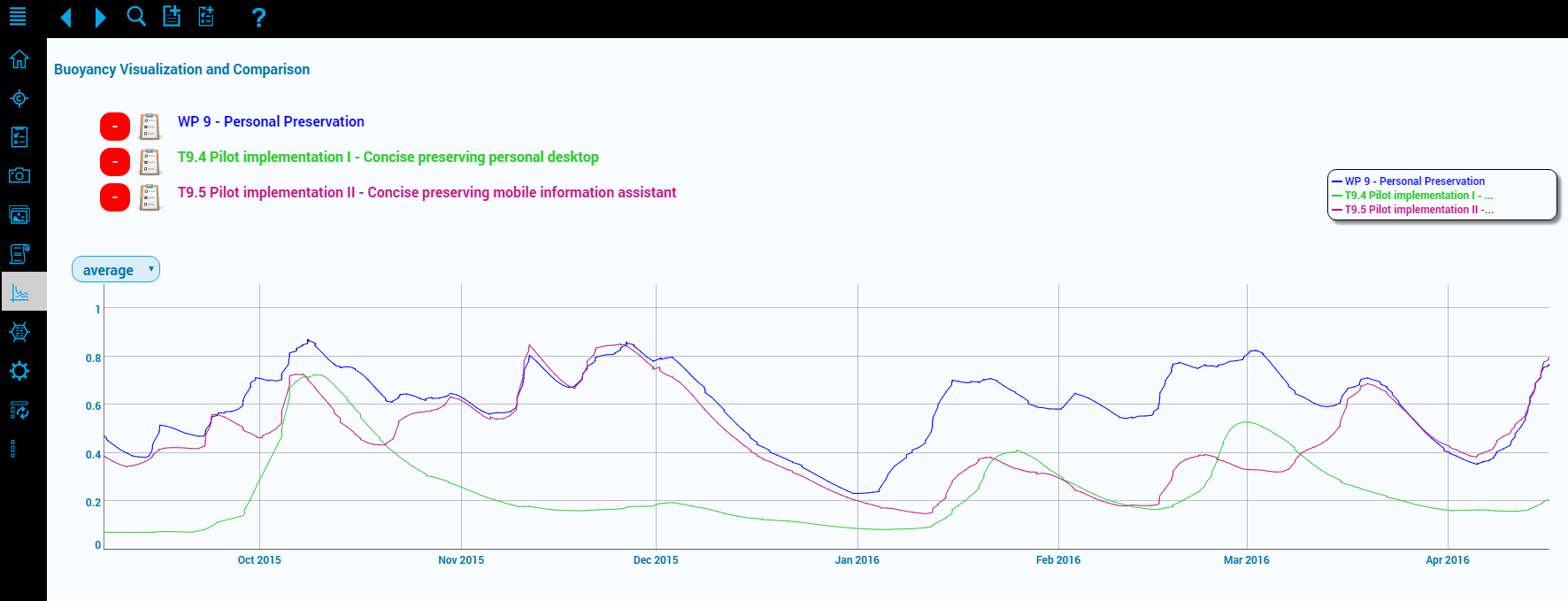}
  \caption{MB graph for a task (blue) and two subtasks (green, pink)}
  \label{fig:mbgraph}
\end{figure}

The major success story we experienced is that rising and fading out of information items works quite well and is already helpful.
Figure \ref{fig:fadingout} shows the example of a workshop:
Viewing this calendar entry during the meeting is depicted on the left-hand side.
All involved persons (top), emails (middle) and tasks (bottom) associated with the event are shown.
After eight months (middle part of the figure), tasks and emails have already dropped in relevance and are thus not shown anymore.
What remains are primarily persons involved in the project as well as the presentation slides and the project thing itself.
After two years (right-hand side), the presentation is not visible anymore (possibly since it was just one presentation among many in the project), whereas photos taken at the event (probably often referred to and shown to colleagues) as well as the project itself remain.
The same is true for several of the involved persons.
A second example especially also showing MB rising is illustrated in Figure \ref{fig:mbgraph}.
It shows the MB graph of a task and two subtasks.
The old task T9.4 (green) is revisited, thus rising in relevance (left-hand side), when writing a deliverable in task T9.5 (pink) and during the final project review (right).
The parent work package task (WP9) is plotted in blue.
\\

\begin{figure}[h!]
  \centering
  \includegraphics[width=1\columnwidth]{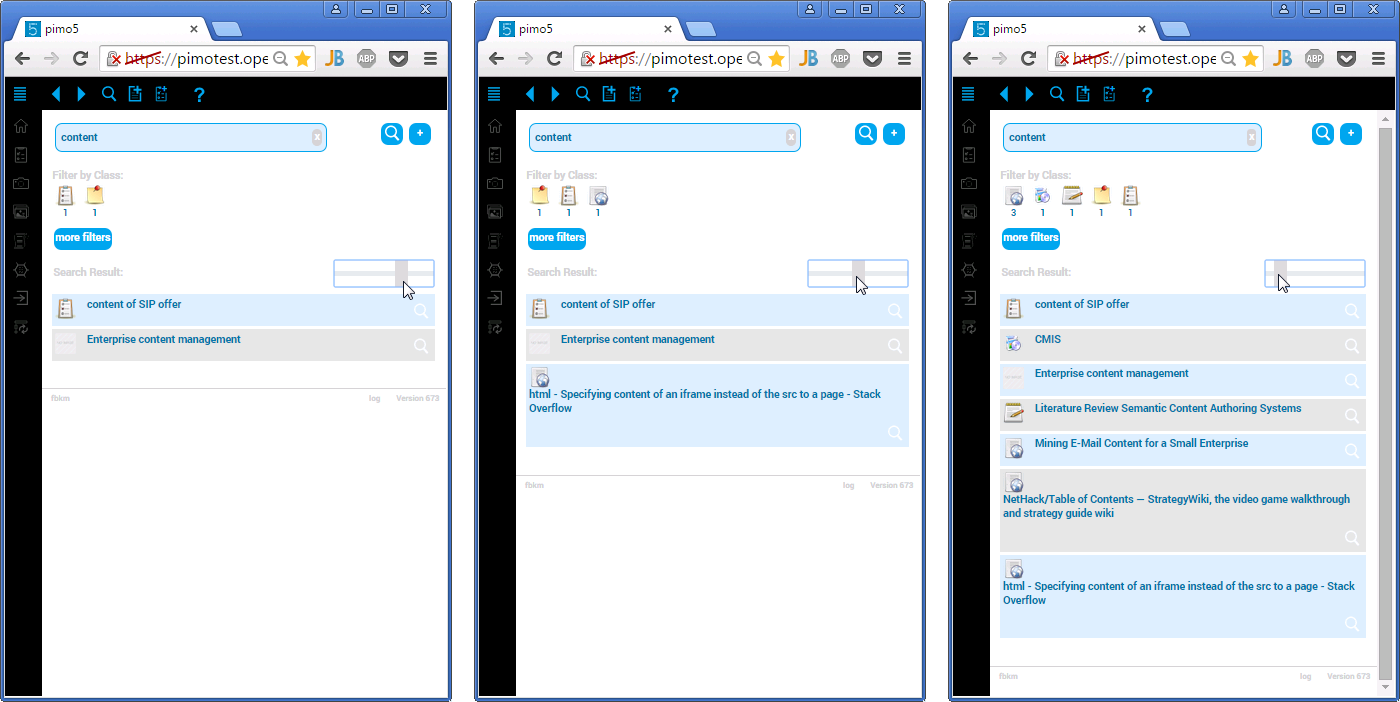}
  \caption{Pathological behaviour in search: habitually lowering the MB threshold instead of rephrasing the query if things are not found in the first place}
  \label{fig:mbslider}
\end{figure}

\noindent
\textbf{Failure Stories \& Problems.}
A failure story we experienced was pathological behavior with our search user interface.
The interface is equipped with a slider to manually lower the MB threshold, which was intended to be used if a user explicitly wants to remember something that has already been forgotten by the system.
Such things have a low buoyancy and would thus not appear in the list of search results otherwise.
What actually happened is depicted in Figure \ref{fig:mbslider} (images left to right):
users started to habitually drag the slider in the low MB direction instead of rephrasing their query if sought things were not found in the first place.
We address this problem in our improved \textit{Forgetful Search} interface, for example by showing users how much of their information is currently covered by the search results.
Additionally, we work on thematically clustering potential search results in the forgotten part so that users get a brief overview where to continue the search if sought things are not yet appearing in the results.
This is addressed in a future paper.

\begin{figure}[t]
  \centering
  \includegraphics[width=1\columnwidth]{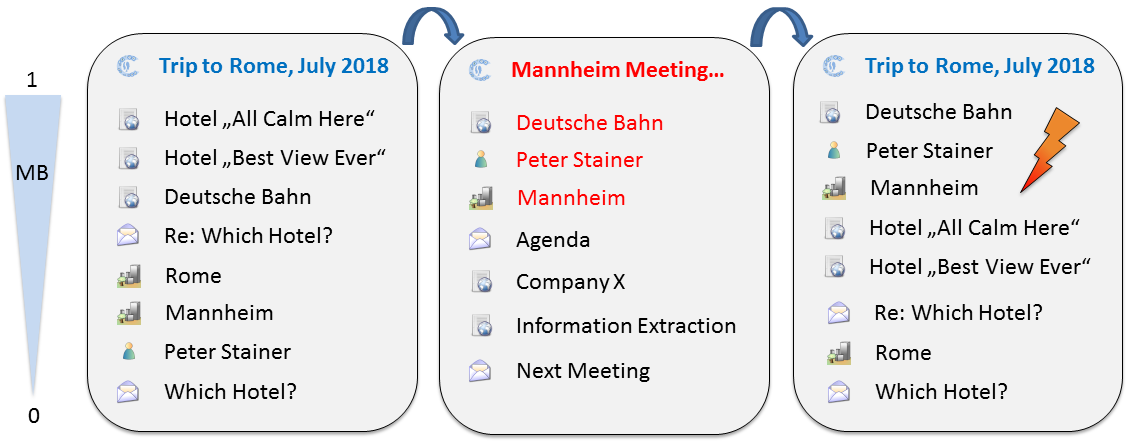}
  \caption{Problem of context-free MB when switching contexts}
  \label{fig:contexts}
\end{figure}

The MB slider failure is not the only problem we experienced:
In \cite{JilekRungeNiederee+2018}, we name more problems and challenges also addressing forgetful information systems in general.
Several issues regarding MB can be traced back to two major problems:
Our initial MB version is too much dependent on the connectivity of the semantic graph (PIMO) and it does not take user context into account.
Let us consider an example to illustrate the latter:
Figure \ref{fig:contexts} (left-hand side) shows the context of a \textit{Trip to Rome in July 2018}.
Its elements are ordered by descending buoyancy.
Looking at these elements from low to high MB, the context's ``golden thread'' could be as follows:
The owner of the context and a colleague, Peter Stainer, planned to meet in Mannheim and then travel on to Rome.
An email was written asking in which hotel to stay.
This email was answered a bit later.
Then, the owner of the context booked a ticket with Deutsche Bahn, checked two different hotels and presumably booked one of them.
At some point, the user switches to another context to plan a different meeting (middle section of Fig. \ref{fig:contexts}).
By chance, Peter Stainer is also an attendee of that meeting, the location is Mannheim and they again travel by train (Deutsche Bahn).
Thus, elements also present in the first context are now stimulated.
When later revisiting the \textit{Trip to Rome} context (Fig. \ref{fig:contexts}, right-hand side), a context-free MB (like our initial version) would destroy the context's ``golden thread'' by re-ordering the elements although there was no action in that particular context which would justify this.

By ``injecting'' contexts into the semantic graph, we are able to kill two birds with one stone:
Working with(in) context spaces, like proposed and realized in \cite{JilekSchroederSchwarz+2018}, ensures that information items are usually associated with at least one context leading to an increase of PIMO's connectivity.
Second, we have a better handling of context switches avoiding cases like the example described before.
Thus, one of the major enhancements of a successor version of MB was to take user context into account, which is discussed in the next section.
\section{Advanced Memory Buoyancy}
\label{sec:advmb}
We learned from cognitive psychology as well as experience in real-world scenarios with MB that our previous research on \textit{context} (e.g. \cite{Schwarz2010,MausSchwarzHaas+2011,JilekMausSchwarz+2015}) can be beneficial for MF.
Thus, the major enhancement we had to incorporate into a successor of our initial MB version was to take user context into account.
This especially also included context switches:
As an addition to the aforementioned gradually changing MB, we needed another value applicable in scenarios of sudden changes.
\\

\begin{figure}[h]
  \centering
  \includegraphics[width=0.93\columnwidth]{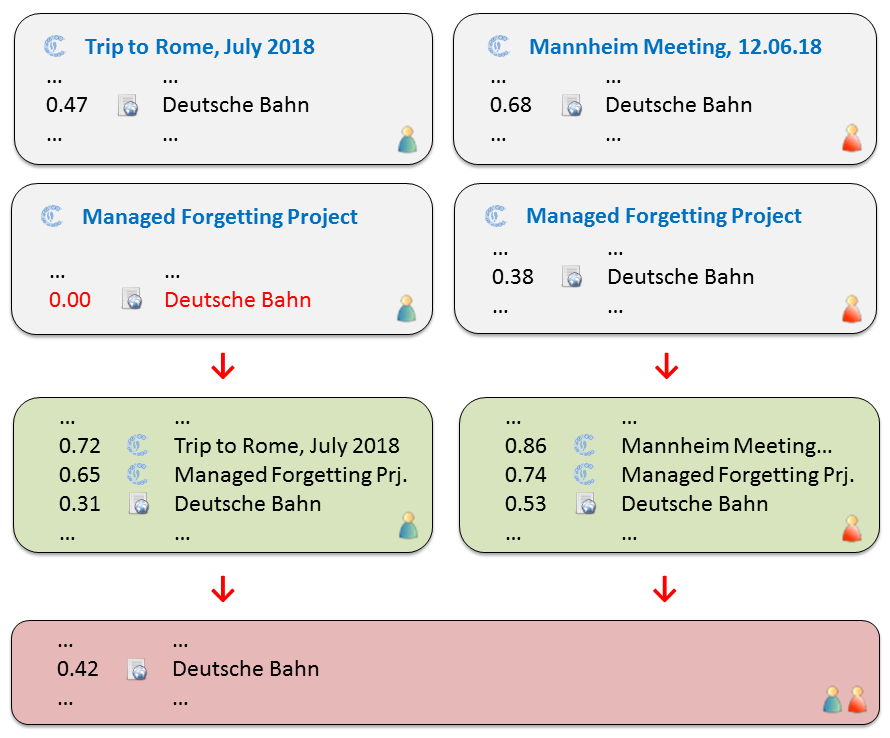}
  \caption{Advanced MB versions: local (top), global (middle), group (bottom)}
  \label{fig:advmb}
\end{figure}

\noindent
\textbf{Local MB.}
We therefore introduced a \textit{local MB}, representing a resource's current value for a specific user in a specific context.
Stated more formally, this results in:
\begin{center}
$f_{\text{MB}_{\textit{local}}}$: ( resource, user, context ) $\rightarrow \text{MB}_{\textit{local}}$
\end{center}
Figure \ref{fig:advmb} (top section, gray shapes) shows different contexts of two users (green: User1, red: User2).
As an example, we see that the website of Deutsche Bahn is relevant for User2 in the context of a Mannheim meeting as well as the Managed Forgetting project.
Its MB varies in both contexts: 0.68 in the first and 0.38 in the second.
The same website is also relevant for User1 in the context of a Trip to Rome.
User1 is also involved in the Managed Forgetting context, a shared context, but for them the Deutsche Bahn website is not relevant there, indicated by an MB value of zero.
This is an example to illustrate that the very same thing can be very important in one context while being totally irrelevant in another.
Second, a certain thing can belong to a certain context (like the Managed Forgetting project) whereas its relevancy in that context varies from user to user (here: 0.00 for User1 vs. 0.38 for User2).
\\

\noindent
\textbf{Global MB.}
We saw that the MB of a certain resource may vary from one context to another.
For several applications (e.g. synchronization, archival) it is helpful to be able to give a summarizing statement about a resource's relevance.
Thus, we also introduced the \textit{global MB} providing a context-free relevancy information of a resource for a certain user:
\begin{center}
$f_{\text{MB}_{\textit{global}}}$: ( resource, user ) $\rightarrow \text{MB}_{\textit{global}}$
\end{center}
It is very similar to our initial MB version and addresses less abrupt changes.
In Figure \ref{fig:advmb}, the green shapes in the middle section contain global MB values.
Note that the individual contexts have their global MB value.
So, when leaving a certain context, the local MB values are frozen and will not change anymore until the context is revisited and therefore active again.
Nevertheless, the MB of the context as a whole, i.e. its global MB, may rise or drop.
In our current prototype, the global and local MB values are calculated independently of each other.
We though plan to investigate the possibility of deriving the global MB from all local values instead of assessing all user evidences multiple times (once for each MB type).
The same is true for the group MB.
\\

\noindent
\textbf{Group MB.}
To especially address our organizational memory scenario, we additionally introduced a \textit{group MB} value representing a resource's relevance for all users of the GIMO (red shape at the bottom of Figure \ref{fig:advmb}):
\begin{center}
$f_{\text{MB}_{\textit{group}}}$: resource $\rightarrow \text{MB}_{\textit{group}}$
\end{center}

\begin{figure}[h]
  \centering
  \includegraphics[width=1\columnwidth]{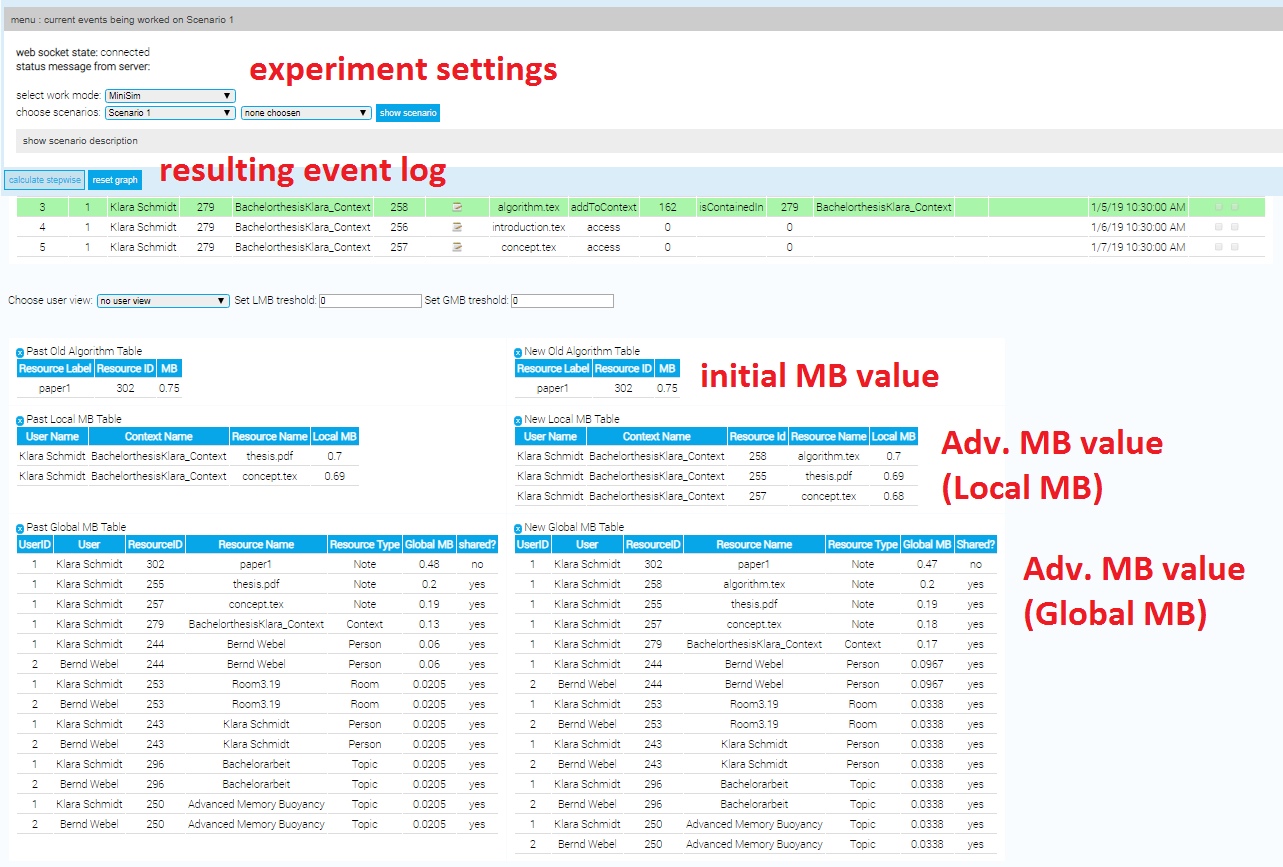}
  \caption{Prototype implementation of Advanced MB for experiments}
  \label{fig:prototype}
\end{figure}

\noindent
\textbf{Experimental Prototype.}
All these advanced MB values are still experimental, especially the group ones.
Since further experiments are necessary, we developed a prototypical implementation depicted in Figure \ref{fig:prototype}.
We have modeled typical knowledge work scenarios:
a single person working on a task, a group of people collaboratively working on a task, whereas in some scenarios only some members of the group are active while others are just readers, a situation shortly before or after an event, etc.
In the first section of the experiment app, one can choose a specific scenario.
Each scenario is represented by a specific list of events, i.e. a generated list of user and system events, serving as input to the MB calculation (together with the semantic graph, PIMO, as the second input source).
The lower half of the figure shows the different MB values: the initial MB as well as the new advanced MB values (group MB not visible on the screenshot).
The left tables contain the MB values before, the right ones the values after the processing of each event.

\section{Conclusion \& Outlook}
\label{sec:conclusion}
In summary, we incorporated MB into our productively used SD-based organizational memory system.
We thus gained three years of practical experience with MB in daily work.
Lessons learned from this as well as further cooperation with cognitive psychologists ultimately led to an advanced MB in the form of local, global and group values.
They were presented in this paper as well as a first prototype to conduct experiments with them.

In the future, we plan to do more user experiments for in-detail evaluation.
We are also interested in scaling experiments using ``big personal data'', i.e. all files of a person's or department's computer(s) instead of a selection stored in an organizational memory.
In \cite{JilekRungeNiederee+2018} we presented more challenges, especially also addressing forgetful information systems in general.
One of the them is gaining and keeping user's trust in such a system, which we also mentioned in this paper when presenting our experience with the misused MB slider (see Section \ref{sec:mb}).
We will especially address this aspect in our future paper on \textit{Forgetful Search}.
\\

\noindent
\textbf{Acknowledgements.}
Parts of this work were funded by the Deutsche Forschungsgemeinschaft (DFG, German Research Foundation) -- DE 420/19-1.

\bibliographystyle{splncs04}
\bibliography{content/refs}

\end{document}